\documentclass[runningheads]{llncs}
\usepackage{subcaption}
\usepackage{amsmath}
\usepackage{amssymb}
\usepackage{hyperref}
\usepackage{multirow}
\usepackage{booktabs}
\usepackage{float}
\usepackage{multicol}
\usepackage{enumitem}
\usepackage{soul}
\usepackage{amsfonts}
\usepackage{CJKutf8}
\usepackage[T1]{fontenc}
\usepackage{breakcites}
\usepackage{graphicx}
\usepackage{amsmath}
\usepackage{ulem}
\usepackage{algpseudocode}
\usepackage{algorithm}
\usepackage[misc]{ifsym}
\begin{document}
\title{MDAP: A Multi-view Disentangled and Adaptive Preference Learning Framework for Cross-Domain Recommendation}
\titlerunning{Multi-view Disentangled and Adaptive Preference Learning for CDR}
%
\author{Junxiong Tong\inst{1,2} \and
Mingjia Yin\inst{1,2} \and
Hao Wang \inst{1,2} \textsuperscript{(\Letter )}\and
Qiushi Pan\inst{1,2} \and
Defu Lian\inst{1,2} \and Enhong chen\inst{1,2} }
\authorrunning{J. Tong et al.}
%
\institute{University of Science and Technology of China, Jinzhai Road,
Hefei, Anhui, China \\
\email{\{tongjunxiong, mingjia-yin, pqs\}@mail.ustc.edu.cn \\ \{wanghao3, liandefu, cheneh\}@ustc.edu.cn} 
\and
State Key Laboratory of Cognitive Intelligence, Hefei, Anhui, China
}
\maketitle              

\begin{abstract}
Traditional recommendation systems, relying on single-domain data, often struggle with sparse data or new user scenarios. Cross-domain Recommendation (CDR) systems leverage multi-domain user interactions to improve performance, especially in sparse data or new user scenarios. However, CDR faces challenges such as effectively capturing user preferences and avoiding negative transfer. To address these issues, we propose the \textbf{\underline{M}ulti-view \underline{D}isentangled and \underline{A}daptive \underline{P}reference Learning (MDAP)} framework. Our MDAP framework uses a multi-view encoder to capture diverse user preferences. The framework includes a gated decoder that adaptively combines embeddings from different views to generate a comprehensive user representation. By disentangling representations and allowing adaptive feature selection, our model enhances recommendations' adaptability and effectiveness. Extensive experiments on benchmark datasets demonstrate that our method significantly outperforms state-of-the-art CDR and single-domain models, providing more accurate recommendations and deeper insights into user behavior across different domains. Our code is available at \url{https://github.com/The-garden-of-sinner/MDAP}.

\end{abstract}

\thispagestyle{empty}
\section{Introduction}

Recommendation systems are essential in modern information retrieval, designed to suggest content based on users' historical behavior and preferences. One of the significant advancements in this field is representation learning, which extracts feature representations from data, thereby enhancing recommendation accuracy and robustness by mapping users and items into latent spaces \cite{Bengio2013}. The integration of deep learning methods, including deep collaborative filtering and graph neural networks, has further improved the performance of recommendation systems by capturing complex and high-level features \cite{He2017, Wang2019}.

Despite these advancements, traditional systems typically rely on single-domain data, limiting their effectiveness in scenarios with sparse data or new users. Cross-domain Recommendation (CDR) systems aim to leverage user interaction data from multiple domains to enhance recommendation performance \cite{Fernandez2012}. By utilizing data from multiple domains, CDR can alleviate the data sparsity issue in a single domain. For example, if a user has limited interaction data in one domain but extensive data in another, CDR can combine this cross-domain data to build a more comprehensive user representation, thereby improving recommendation accuracy and robustness \cite{Man2017}. However, CDR systems face several significant challenges:

\begin{itemize}
    \item \textbf{Preference Heterogeneity:} 
    The preference heterogeneity in cross-domain user behavior is a major challenge. User behaviors in different domains may be driven by distinct latent preferences. For instance, a user might prioritize the genre when reading novels but consider the director first when selecting movies. These preference features are often entangled within a single domain, making it difficult to generalize behavior from one domain to another. In CDR, addressing this preference heterogeneity is crucial to more accurately capture user preferences across different domains.

    \item \textbf{Feature Space Disparity:} 
    Different domains may have significantly different feature spaces, making direct transfer learning less effective and potentially causing negative transfer. Negative transfer occurs when transfer learning not only fails to improve performance but also degrades the model's performance in the target domain. Effective methods are needed to handle these disparities and achieve successful cross-domain recommendations.
\end{itemize}


To address the above issues, we propose a novel framework called Multi-View Disentangled and Adaptive Preference Learning (MDAP). By disentangling multi-view encoding and adaptive feature selection, we aim to better represent user behavior and capture diverse user preferences across different domains.

Specifically, our approach employs a multi-view encoder that uses Gumbel-Softmax techniques to generate multiple soft assignment matrices, capturing various aspects of user preferences. These matrices guide the encoding of user-item interaction data into multiple embeddings, each representing a different perspective of user preferences. The framework further integrates a gated decoder where domain-specific gating networks adaptively combine embeddings from different perspectives and decode the combined embeddings to produce comprehensive user representations. By disentangling representations and allowing adaptive feature selection, our model enhances the adaptability and effectiveness of recommendations.

The primary contributions of our work can be summarized as follows:

\begin{itemize}
    \item \textbf{Addressing Preference Heterogeneity:} 
    We decouple and select multi-view user preferences to mitigate negative transfer, capturing diverse user preferences across domains with a multi-view encoder.

    \item \textbf{Proposing a Multi-View Encoder:} 
    Using Gumbel-Softmax techniques, our encoder generates multiple soft assignment matrices, capturing various aspects of user preferences and enhancing system adaptability.

    \item \textbf{Designing Domain-Specific Gating Mechanisms:} 
    Our gated decoder, with domain-specific gating networks, adaptively combines embeddings from different views, ensuring effective decoupling and accurate recommendation matching.

    \item \textbf{Implementing Knowledge Transfer:} 
    By sharing encoders and decoders across domains, we achieve effective knowledge transfer, enhancing overall system performance despite domain differences.

    \item \textbf{Extensive Experimental Validation:} 
    Experiments on benchmark datasets show our method significantly outperforms state-of-the-art CDR and single-domain models, providing more accurate recommendations and deeper user insights.
\end{itemize}

The remainder of this paper is organized as follows. In Section \ref{related work}, we review related work in the fields of recommendation systems and cross-domain recommendation. Section \ref{method} presents the details of our proposed MDAP framework. In Section \ref{experiment}, we describe the experimental setup and results. Finally, we conclude the paper and discuss future work in Section \ref{conclusion}.

\section{Related work}\label{related work}
In this section, we review the literature on traditional recommendation systems and cross-domain recommendation approaches, with a particular focus on methodologies pertinent to dual-target scenarios.

\subsection{Single Domain Recommendation}
Traditional recommendation systems have extensively utilized matrix factorization techniques to map users and items into a latent space, capturing interaction preferences. Bayesian Personalized Ranking (BPR)~\cite{rendle2012bpr} optimizes pair-wise ranking loss.
With the rise of deep learning, nonlinear models such as Neural Matrix Factorization (NeuMF) have demonstrated superior performance over linear models.
The graph structure inherent in user-item interaction matrices has inspired numerous Graph Neural Network (GNN)-based recommendations. Noteworthy contributions include NGCF~\cite{wang2019ngcf} and LightGCN~\cite{he2020lightgcn}, which model high-order user-item relationships. Furthermore, UltraGCN~\cite{mao2021ultragcn} approximates infinite-layer graph convolutions for simplification, while PTGCN~\cite{PTGCN} effectively integrates temporal dynamics for session-based recommendations. SCG-SPRe~\cite{SCG-SPRe} leverages substitutable-complementary graphs for sequential recommendation, capturing temporal patterns in behavior sequences.
Recent self-supervised contrastive learning methods, such as SGL~\cite{wu2021sgl} and SimGCL~\cite{yu2022simgcl}, have shown notable performance improvements by leveraging graph contrastive learning techniques. DirectAU~\cite{wang2022directau} achieves impressive results through direct representation alignment and uniformity constraints.
Large language models(LLMs) are also widely applied in recommendation systems, demonstrating significant performance and flexibility~\cite{wu2024survey,yin2024entropy,shen2024exploring}.In addition, Sequential Recommendation represents a significant area of research within the field of recommendation systems, where methods such as feature decoupling~\cite{han2024efficient,han2024end4rec}, unsupervised learning~\cite{han2023guesr}, diffusion models~\cite{xie2024bridging}, and adaptive techniques~\cite{yin2023apgl4sr} have been applied, demonstrating promising results.

\subsection{Dual-Target Cross-Domain Recommendation}
Cross-domain recommendation (CDR) methods aim to alleviate data sparsity and cold-start problems by leveraging richer data from related domains. This paper focuses on dual-target CDR, aiming to concurrently enhance performance across two domains by leveraging data from both.

Several studies have explored transferring user preferences across domains. Collective Matrix Factorization (CMF)~\cite{singh2008cmf} and Cross-Domain Factorization Machines (CDFM)~\cite{loni2014crossfm} extend traditional matrix factorization and FM models by incorporating transfer modules for cross-domain adaptation of user representations.
With the advent of deep learning, MLP-based approaches such as CoNet~\cite{hu2018conet}, DTCDR~\cite{zhu2019dtcdr}, and GA-DTCDR~\cite{zhu2020gadtcdr} have introduced various neural network-based knowledge transfer modules, including cross-connection networks and feature combination strategies. DDTCDR~\cite{li2020ddtcdr} further incorporates a latent orthogonal mapping function to enhance domain adaptation. ETL~\cite{chen2023feature_suppresion} proposes preference transfer based on equivalent transformations, modeling joint distributions across domains.
Graph neural networks have significantly contributed to CDR; PPGN~\cite{zhao2019ppgn} applies graph convolution on an integrated graph combining two domains, while Bi-TGCF~\cite{liu2020cross} employs domain-specific GCN modules alongside a feature transfer module within each graph layer to facilitate knowledge sharing among common users. AADA~\cite{AADA} connects cold-start and warm-start users through a hypergraph approach, aiming to bridge the gap between users.

To alleviate negative transfer, recent works like DR-MTCDR~\cite{guo2022drmtcdr}, CDRIB~\cite{cao2022cdrib}, and DisenCDR~\cite{cao2022disencdr} explicitly model domain-common and domain-specific representations using techniques such as graph contrastive learning, information bottleneck methods, and mutual information-based disentanglement objectives.In contrast, our proposed MDAP framework differentiates itself by adopting a multi-view perspective. Specifically, MDAP integrates multiple views of user interactions across different domains to capture nuanced and fine-grained patterns that are often overlooked by traditional single-view approaches.This work has also been inspired by some other studies~\cite{wang2021hypersorec,wang2019mcne} within our lab.  

This comprehensive review lays the groundwork for our proposed MDAP framework by highlighting key advancements in both single-domain and cross-domain recommendation systems.
We will next present formal definitions for the dual-target cross-domain recommendation problem.

\section{Methodology}\label{method}



\begin{figure*}[htbp]
    \centering
    \includegraphics[width=\textwidth]{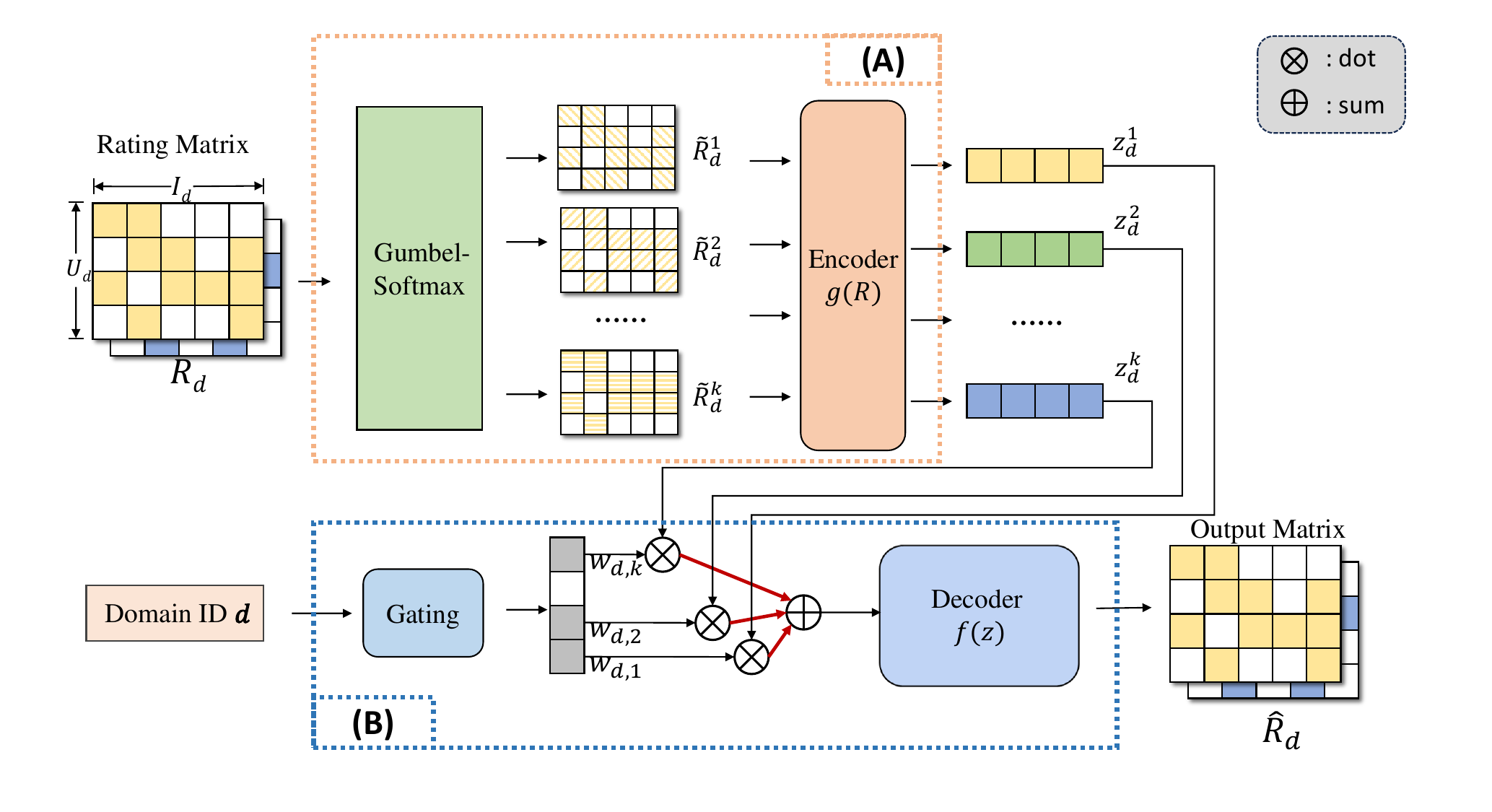}
    \caption{The Framework of \textbf{MDAP}: (A) The \textbf{Multi-view Preference Encoder} (orange dashed box) employs Gumbel-Softmax to generate diverse user preference embeddings from the rating matrix. (B) The \textbf{Domain-specific Gated Decoder} (blue dashed box) uses domain-specific gated networks to combine these embeddings based on the domain ID, forming a unified user representation that is then decoded to reconstruct the user-item interaction matrix.}
    \label{fig:mdap_framework}
\end{figure*}

In this section, we first describe the problem definition of this paper(Section \ref{problem def}), and then introduce the details of our proposed MDAP framework, as shown in Figure \ref{fig:mdap_framework}. Our approach employs a multi-view preference encoder using the Gumbel-Softmax technique to generate multiple soft assignment matrices, capturing diverse aspects of user preferences. These matrices guide the encoding of user-item interaction data into multiple embeddings, each representing a different view of user preferences (Section \ref{gumbel softmax}). Then our framework incorporates a domain-specific gated decoder, which adaptively combines the embeddings from different views. The combined embeddings are decoded to produce the final recommendation scores of users in both domains (Section \ref{gated}). Finally, the overall objective function is formulated (Section \ref{objective}), and the training process is stated in pseudo-code (Algorithm \ref{alg:training_process}).

\subsection{Problem Definition}\label{problem def}

Cross-domain recommendation (CDR) aims to leverage information from multiple domains to enhance the recommendation performance in both the source and target domains. Formally, consider two domains: a source domain $\mathcal{S}$ and a target domain $\mathcal{T}$. Each domain has a set of users, with $\mathcal{U}_\mathcal{S}$ in the source domain and $\mathcal{U}_\mathcal{T}$ in the target domain. The set of overlapping users is denoted as $\mathcal{U}_{\mathcal{S} \cap \mathcal{T}} = \mathcal{U}_\mathcal{S} \cap \mathcal{U}_\mathcal{T}$, and the non-overlapping users are $\mathcal{U}_\mathcal{S} \setminus \mathcal{U}_\mathcal{T}$ and $\mathcal{U}_\mathcal{T} \setminus \mathcal{U}_\mathcal{S}$. The items in the source and target domains are $\mathcal{I}_\mathcal{S}$ and $\mathcal{I}_\mathcal{T}$, respectively. The user-item interaction matrices for the source and target domains are denoted as $\mathbf{R}_\mathcal{S} \in \mathbb{R}^{|\mathcal{U}_\mathcal{S}| \times |\mathcal{I}_\mathcal{S}|}$ and $\mathbf{R}_\mathcal{T} \in \mathbb{R}^{|\mathcal{U}_\mathcal{T}| \times |\mathcal{I}_\mathcal{T}|}$, respectively, where $r_{ui}$ represents the interaction (e.g., rating or implicit feedback) between user $u$ and item $i$. In our framework, notably, we extend both interaction matrices by padding with zeros, and then obtain $\mathbf{R}_\mathcal{S} \in \mathbb{R}^{|\mathcal{U}| \times |\mathcal{I}_\mathcal{S}|}$ and $\mathbf{R}_\mathcal{T} \in \mathbb{R}^{|\mathcal{U}| \times |\mathcal{I}_\mathcal{T}|}$, where $\mathcal{U}$ denotes the set of unique users across both domains.

The goal of CDR is to predict the interaction $\hat{r}_{uj}$ between user $u$ and item $j$ in both the source domain $\mathcal{S}$ and the target domain $\mathcal{T}$ by leveraging interactions from both domains.

\subsection{Multi-view Preference Encoder}\label{gumbel softmax}

The multi-view preference encoder is designed to encode users into \(k\) distinct embeddings, with each embedding representing a unique perspective of user preferences. To achieve this, we first generate category assignment logits, denoted as \(\mathbf{C} \in \mathbb{R}^{|\mathcal{U}| \times k}\), where \(k\) is the number of different views. The generation of category assignment logits \(\mathbf{C}\) involves several steps: initially, the item embeddings and core embeddings are normalized. Subsequently, the interaction matrix is normalized and dropout is applied. Finally, the category assignment logits \(\mathbf{C}\) are produced through matrix multiplication of the normalized item embeddings and core embeddings. These logits \(\mathbf{C}\) are then utilized by the decoder to generate \(k\) user embeddings, each capturing a different aspect of user preferences.

Next, the Gumbel-Softmax technique is used to generate \(k\) soft assignment matrices from the category logits matrix. These matrices are used to weight the user-item interaction matrix, producing view-specific inputs. These inputs are then encoded to obtain \(k\) different view-specific embeddings for each user.

Given the user-item interaction matrix \(\mathbf{R}_d\) and the category logits matrix \(\mathbf{C}\), the specific steps are as follows:

\begin{equation}
\mathbf{G}_{ij} = -\log(-\log(U_{ij})), \quad U_{ij} \sim \text{Uniform}(0, 1)\label{noise}
\end{equation}
\begin{equation}
\mathbf{S}_{ij} = \frac{\exp((\log(\mathbf{C}_{ij}) + \mathbf{G}_{ij}) / \tau)}{\sum{j=1}^{k} \exp((\log(\mathbf{C}_{ij}) + \mathbf{G}_{ij}) / \tau)}\label{matrix_cal}
\end{equation}

Here, \(U_{ij}\) is drawn from a uniform distribution to generate the Gumbel noise \(\mathbf{G}\). The soft assignment matrix \(\mathbf{S}\) is computed by applying the Gumbel-Softmax function to the logits matrix \(\mathbf{C}\), where \(\tau\) is the temperature parameter.

The interaction matrix \(\mathbf{R}_d\) is weighted by \(\mathbf{S}\) to produce view-specific inputs:
\begin{equation}
\mathbf{\tilde{R}}_d^i = \mathbf{R}_d \odot \mathbf{S}^i \quad \text{for} \ i = 1, 2, \ldots, k
\label{rdi}
\end{equation}
where \(\odot\) denotes element-wise multiplication. 

By this means, we obtain user-item interaction matrices of multiple views. The diversity among the views is enforced through the use of Gumbel-Softmax, which allows for differentiable sampling of discrete distributions. These inputs are then encoded to embeddings, formally:
\begin{equation}
\mathbf{E}_d^i = g(\mathbf{\tilde{R}}_d^i)
\label{edi}
\end{equation}
where \(g\) is typically a multi-layer perceptron (MLP), and \(\mathbf{E}_d^i \in \mathbb{R}^{m \times l}\) is the embedding matrix for the \(i\)-th view.

\subsection{Domain-specific Gated Decoder}\label{gated}

After encoding users to multi-view preference embeddings, the gated decoder integrates them into a unified user representation in a domain-specific way. This integration is controlled by domain-specific gated networks that dynamically adjust the contribution of each view.

Let $\mathbf{w}_d \in \mathbb{R}^{k}$ be the gating weights for domain $d$, where $k$ is the number of different views. These weights determine the importance of each view in forming the combined user embedding for that domain. The combined embedding $\mathbf{z}_d$ is calculated as follows:
\begin{equation}
\mathbf{z}_d = \sum_{i=1}^{k} w_{d,i} \mathbf{E}_d^i
\label{zd}
\end{equation}
where $\mathbf{E}_d^i$ denotes the $i$-th view-specific user embedding for domain $d$, and $w_{d,i}$ is the corresponding gating weight.

The gating weights $\mathbf{w}_d$ are generated by a domain-specific gated network. The input to this network is the domain ID $d$, which is mapped to the weight vector $\mathbf{w}_d$ through a single-layer embedding. The process is as follows:
\begin{equation}
{w}_d = \text{softmax}(\mathbf{W}_e[d])
\label{gate}
\end{equation}
where $\mathbf{W}_e$ is the embedding matrix, $\mathbf{W}_e[d]$ denotes the embedding vector corresponding to domain $d$, and the $\text{softmax}$ function ensures the weights are normalized to sum to 1.

The gating mechanism achieves decoupling primarily by combining the learning of domain-specific gating and imposing orthogonality constraints on the gating for different domains. By learning different gates for each domain, the model can capture and utilize features unique to each domain, thereby enhancing its ability to adapt to diverse domains. Simultaneously, imposing orthogonality constraints on the embedding vectors ensures that the embeddings for different domains are orthogonal to each other. This promotes independence among the learned representations and reduces the risk of negative transfer between domains. Consequently, the model can maintain domain-specific performance while achieving better generalization across multiple domains.

Once the combined embedding $\mathbf{z}_d$ is obtained, the decoder reconstructs the user-item interaction matrix $\hat{\mathbf{R}}_d$ for domain $d$. This reconstruction is typically achieved using a multi-layer perceptron (MLP), denoted as $f$:
\begin{equation}
\hat{\mathbf{R}}_d = f(\mathbf{z}_d)
\label{hrd}
\end{equation}

The gated network is trained based on domain-specific features to generate the weights $\mathbf{w}_d$. This training ensures that the gated network effectively combines the different view-specific embeddings in a way that is optimal for each domain. By doing so, the model enhances its adaptability and effectiveness across different domains, resulting in more accurate and personalized user representations.

\subsection{Objective Function}\label{objective}

The overall objective function of the model is to minimize the reconstruction loss while ensuring effective disentanglement of user representations across domains. This can be formalized as:

\begin{equation}
\mathcal{L} = \sum_{d} \left( \| \mathbf{R}_d - \hat{\mathbf{R}}_d \|^2 \right) + \lambda ({w}_{\mathcal{S}} \cdot {w}_{\mathcal{T}})\label{loss_cal}
\end{equation}

where \(\mathbf{R}_d\) is the original user-item interaction matrix for domain \(d\), \(\hat{\mathbf{R}}_d\) is the reconstructed interaction matrix, and \(\lambda\) denotes a hyperparameter controlling the regularization strength.

The training process for the proposed model is outlined in Algorithm \ref{alg:training_process}. The model parameters, including the encoder and decoder weights, as well as the gating network parameters, are updated iteratively to minimize the loss function.


\begin{algorithm}[t]
\caption{MDAP Algorithm}
\label{alg:training_process}
\begin{algorithmic}[1]
\Require User-item interaction matrices \( \mathbf{R}_d \) for each domain \( d \), number of views \( k \), temperature parameter \( \tau \), number of epochs \( N \), learning rate \( \eta \)
\Ensure Trained model parameters

\State \textbf{Initialization:} Initialize category logits matrix \( \mathbf{C} \) and model parameters

\For{epoch = 1 to \( N \)}
    \For{each domain \( d \)}
        \For{each user-item pair \( (u, v) \)}
             \State Sample Gumbel noise: $G_{ij}$ by Eq.(\ref{noise})
            \State Compute soft assignment matrix: $S_{ij}$ by Eq.(\ref{matrix_cal})
        \EndFor
        \For{each view \( i \)}
            \State Generate view-specific input: $\mathbf{\tilde{R}}_d^i$ by Eq.(\ref{rdi})
            \State Encode to obtain embeddings: $\mathbf{E}_d^i$ by Eq.(\ref{edi})
        \EndFor
    \EndFor
    \For{each domain \( d \)}
        \State Compute gating weights ${w}_d$ by Eq.(\ref{gate})
        
        \State Compute combined embedding: $\mathbf{z}_d$ by Eq.(\ref{zd})
        \State Decode to produce recommendation scores: $\hat{\mathbf{R}}_d $ by Eq.(\ref{hrd})
    \EndFor
    \State Compute loss function \( \mathcal{L} \) by Eq.(\ref{loss_cal})
    \State Update model parameters using gradient descent with learning rate \( \eta \)
\EndFor

\end{algorithmic}
\end{algorithm}

\section{Experiment}\label{experiment}
In this section, we comprehensively evaluate the MDAP framework. Section~\ref{Experimental Settings} outlines the datasets, baseline methods, and experimental setup. Section~\ref{Performance Comparison} demonstrates our approach's superior performance against all baselines. The effectiveness of each module within MDAP is analyzed in Section~\ref{ablation}.

\subsection{Experimental Settings} \label{Experimental Settings}

\subsubsection{\textbf{Datasets}}
We validate our method on three widely adopted datasets:
\begin{itemize}
    \item \textbf{Epinions}\footnote{https://www.cse.msu.edu/~tangjili/datasetcode/truststudy.htm}: This dataset comes from a who-trust-whom online social network of a general consumer review site, and spans 587 domains, capturing user ratings (1-5), reviews, profiles, item features, and trust statements.
    \item \textbf{Douban}\footnote{https://recbole.s3-accelerate.amazonaws.com/CrossDomain/Douban.zip}: This dataset is built from Douban, a popular review site, covering books, movies, and music. It includes user profiles, tag-based preferences, item attributes, and social relationships.
    \item \textbf{Amazon}\footnote{http://snap.stanford.edu/data/amazon/productGraph/categoryFiles/}: This dataset is collected from the Amazon web store, which includes ratings (1-5), review texts, timestamps, and item categories.
\end{itemize}
For pre-processing, we treat different product categories as different domains, selecting two from each dataset: Electronics and Games from Epinions; Book and Movie from Douban; and Electronics and Movie \& TV from Amazon. We filter out users and items with fewer than five interactions for Douban and Amazon.

\subsubsection{\textbf{Compared Baselines}}
We compare MDAP with state-of-the-art methods across three categories: single-domain (SD), multi-task learning (MTL), and cross-domain (CD). For SD methods, we include BPRMF~\cite{rendle2012bpr}, NGCF~\cite{wang2019ngcf}, LightGCN~\cite{he2020lightgcn}, MultiVAE~\cite{liang2018multivae}, and DirectAU~\cite{wang2022directau}. For MTL approaches, we evaluate MMoE~\cite{ma2018mmoe} and PLE~\cite{tang2020ple}. For CD methods, we consider CoNet~\cite{hu2018conet}, DTCDR~\cite{zhu2019dtcdr}, Bi-TGCF~\cite{liu2020cross}, and DR-MTCDR~\cite{guo2022drmtcdr}. Single-domain methods use mixed interaction records from both domains as input.

\subsubsection{\textbf{Evaluation Protocols}}
We partition each dataset into training (80\%), validation (10\%), and test (10\%) sets. To evaluate the performance of top-k recommendations, we utilize Recall@20 and Normalized Discounted Cumulative Gain(NDCG)@20 as evaluation metrics~\cite{he2020lightgcn,wang2022directau}. Using Recall and NDCG together provides a comprehensive evaluation of recommendation systems, balancing the need for completeness with the importance of item ranking, thus guiding system optimization effectively.

\subsubsection{\textbf{Implementation Details}}
All methods are implemented based on RecBole-CDR\cite{zhao2021recbole,zhao2022recbole}. The maximum number of epochs is 1000 with early stopping after 20 epochs of no improvement in NDCG@20. The batch size is 4096, and Adam\cite{kingma2014adam} is the optimizer. The embedding dimension is set to 64. Our method's learning rate is 1e-3. We tune dropout probability, and $\tau$ first, then $k$, and $\lambda$. Baseline hyperparameters are tuned as suggested by their original papers. The implementation code is available on the website.
\begin{table}[h]
\centering
\caption{Hyper-parameter settings for each dataset.}
\label{tab:hyper-parameter}
\begin{tabular}{l|l|lll}
\hline
\textbf{Hyper-parameter} & \textbf{Search Range} & \textbf{Epinions} & \textbf{Douban} & \textbf{Amazon} \\ \hline
Dropout Probability      & \{0.5, 0.7, 0.9\}     & 0.5               & 0.7             & 0.7            \\
Softmax Temperature $\tau$ & \{0.1, 0.2, 0.5\}        & 0.2               &  0.1              & 0.1              \\
Number of Factors $k$    & \{4, 8, 16\}          & 8                 & 16              & 4              \\ 
regularization strength $\lambda$ & \{0.1,0.5,1\}  &0.5                  &0.1             &0.1            \\
\hline
\end{tabular}
\end{table}

\begin{table*}
\centering
\vspace{0.2cm}
  \caption{Recall@20 Results on datasets. Best results are highlighted in bold, with the second-best results underlined. Our model demonstrates consistent superiority over all baselines, achieving statistical significance with p-value < 0.01.}
  \label{tab:overall results}
  \renewcommand{\arraystretch}{1.1}
  \resizebox{0.9\linewidth}{!}{
  \begin{tabular}{|c|c|c|c|c|c|c|c|}
	\hline
        \multicolumn{2}{|c|}{Datasets} & \multicolumn{2}{|c|}{Epinions} & \multicolumn{2}{|c|}{Douban} & \multicolumn{2}{|c|}{Amazon} \\
        \hline
        \multicolumn{2}{|c|}{Domains} & \multicolumn{1}{|c|}{Elec} & \multicolumn{1}{|c|}{Games} & \multicolumn{1}{|c|}{Book} & \multicolumn{1}{|c|}{Movie} & \multicolumn{1}{|c|}{Elec} & \multicolumn{1}{|c|}{Movie} \\
        \hline

       \multirow{5}{*}{SD} & BPR & 0.0280  & 0.1618  & 0.1579  & 0.1508 & 0.0726  & 0.0316 \\
        ~ & NGCF & 0.0412  & 0.1861 & 0.1662 & 0.1697 & 0.0785 & 0.0518 \\
        ~ & LightGCN & 0.0506 & 0.2134 & 0.1840 & 0.1634 & 0.0976 & 0.0470 \\
        ~ & MultVAE & 0.0424 & 0.2035 & 0.2023 & \underline{0.2210} & 0.0918 & 0.0484\\
        ~ & DirectAU & 0.0461 & \underline{0.2409} & \underline{0.2267} & 0.1703 & \underline{0.1049} & \underline{0.0549} \\

        \hline
        \multirow{2}{*}{MTL} & MMoE & 0.0464 & 0.1636 & 0.1233 & 0.1202 & 0.0423 & 0.0441 \\
        ~ & PLE & 0.0543 & 0.1698 & 0.1273 & 0.1255 & 0.0451 & 0.0449 \\

        \hline
        \multirow{4}{*}{CD} & CoNet & 0.0533 & 0.1683 & 0.1075 & 0.1551 & 0.0382 & 0.0445 \\
        ~ & DTCDR & 0.0520 & 0.1716 & 0.1047 & 0.1341 & 0.0366 & 0.0435 \\

        ~ & Bi-TGCF & 0.0534 & 0.2232 & 0.1865 & 0.1900 & 0.0893 & 0.0515 \\
        ~ & DR-MTCDR & \underline{0.0566} & 0.1925 & 0.1509 & 0.1645 & 0.0476 & 0.0423 \\
        \hline

        ~ & MDAP & \textbf{0.0664} & \textbf{0.2416} & \textbf{0.2839} & \textbf{0.2367} & \textbf{0.1276} & \textbf{0.0577} \\
        \hline
  \end{tabular}
  }
\end{table*}

\begin{table*}[ht]
\centering
\vspace{0.2cm}
\caption{NDCG@20 Results on datasets. Best results are highlighted in bold, with the second-best results underlined. Our model demonstrates consistent superiority over all baselines, achieving statistical significance with p-value < 0.01.}
\label{tab:ngcd}
\renewcommand{\arraystretch}{1.1}
\resizebox{0.9\linewidth}{!}{
\begin{tabular}{|c|c|c|c|c|c|c|c|c|}
\hline
\multicolumn{2}{|c|}{Datasets} & \multicolumn{2}{c|}{Epinions} & \multicolumn{2}{c|}{Douban} & \multicolumn{2}{c|}{Amazon} \\
\hline
\multicolumn{2}{|c|}{Domains} & Elec & Games & Book & Movie & Elec & Movie \\
\hline

\multirow{5}{*}{SD} & BPR & 0.0157 & 0.0784 & 0.0741 & 0.1044 & 0.0308 & 0.0148 \\
~ & NGCF & 0.0214 & 0.0843 & 0.0820 & 0.1169 & 0.0343 & 0.0227 \\
~ & LightGCN & 0.0240 & 0.1014 & 0.1017 & 0.1187 & 0.0437 & 0.0223 \\
~ & MultVAE & 0.0221 & 0.0973 & 0.1178 & \underline{0.1537} & 0.0424 & 0.0209 \\
~ & DirectAU & 0.0241 & \underline{0.1179} & \underline{0.1361} & 0.1146 & \underline{0.0517} & \underline{0.0257} \\
\hline

\multirow{2}{*}{MTL} & MMoE & 0.0221 & 0.0689 & 0.0630 & 0.0940 & 0.0172 & 0.0187 \\
~ & PLE & 0.0238 & 0.0703 & 0.0652 & 0.0975 & 0.0192 & 0.0201 \\
\hline

\multirow{4}{*}{CD} & CoNet & 0.0232 & 0.0718 & 0.0467 & 0.1109 & 0.0161 & 0.0185 \\
~ & DTCDR & 0.0236 & 0.0705 & 0.0447 & 0.0968 & 0.0143 & 0.0181 \\
~ & Bi-TGCF & 0.0239 & 0.1073 & 0.0852 & 0.1193 & 0.0333 & 0.0238 \\
~ & DR-MTCDR & \underline{0.0251} & 0.0869 & 0.0715 & 0.1120 & 0.0208 & 0.0193 \\
\hline

~ & MDAP & \textbf{0.0311} & \textbf{0.1248} & \textbf{0.1812} & \textbf{0.1778} & \textbf{0.0628} & \textbf{0.0273} \\
\hline
\end{tabular}
}
\end{table*}

\subsection{Performance Comparison} \label{Performance Comparison}

Table \ref{tab:overall results} and \ref{tab:ngcd} show the performance comparison of all methods on three CDR datasets in terms of Recall@20 and NDCG@20. The experimental results lead to the following key observations: 

\begin{itemize}
    \item On dual-domain datasets, SD methods tend to perform relatively well in one domain while lagging behind in another. This limitation arises because SD methods are typically designed to optimize for domain-specific features and fail to generalize across domains, unlike CDR methods that are inherently designed to transfer knowledge between domains.
    \item Different CDR methods exhibit varying performance across different datasets.  Specifically, CoNet and DTCDR exhibit notably poor performance on the Amazon dataset. Bi-TGCF demonstrates relatively strong performance on the Epinions and Douban datasets but shows mediocre results on the Amazon dataset. Similarly, DR-MTCDR performs well on the Epinions dataset but underperforms on the Amazon dataset. These findings underscore the necessity of designing adaptive transfer modules to address the issue of negative transfer in cross-domain scenarios.
    \item Compared with several representative CDR methods, MDAP has consistently demonstrated its superiority across all three datasets, significantly outperforming every baseline method. These results underscore MDAP's proficiency in capturing intricate combinations of user preferences and identifying the most relevant preference features across diverse domains.
\end{itemize}

\begin{table}
\centering
\vspace{-0.10cm}
  \caption{
  Recommendation Performance Comparison among Different Model Variants of MDAP: MDAP-GS (without Gumbel-Softmax), MDAP-MV (without Multi-View), and MDAP-DG (without Domain-specific Gating)
  }
  \label{tab: ablation study}
  \renewcommand{\arraystretch}{1.1}
  \resizebox{0.9\columnwidth}{!}{
  \begin{tabular}{|c|c|c|c|c|c|c|}
	\hline
	\multicolumn{2}{|c|}{Datasets} &
        Metrics@20 &
        MDAP &
        MDAP-GS &
        MDAP-MV &
        MDAP-DG  \\
        \hline
        \multirow{4}{*}{Epinions} & \multirow{2}{*}{Elec} & Recall & \textbf{0.0664}& 0.0564& 0.0497& 0.0547\\
        ~ & ~ & NDCG & \textbf{0.0311}& 0.0271& 0.0254& 0.0266\\
		\cline{2-7}
        ~ & \multirow{2}{*}{Games} & Recall & \textbf{0.2416}& 0.2322& 0.2300& 0.2294\\
        ~ & ~ & NDCG & \textbf{0.1248}& 0.1154& 0.1164& 0.1111\\

        \hline
        \multirow{4}{*}{Douban} & \multirow{2}{*}{Book} & Recall & \textbf{0.2839} & 0.1632& 0.1731& 0.1686 \\
        ~ & ~ & NDCG & \textbf{0.2412}& 0.0990& 0.1003& 0.1285 \\
		\cline{2-7}
        ~ & \multirow{2}{*}{Movie} & Recall & \textbf{0.2367}& 0.2320& 0.2354& 0.2153\\
        ~ & ~ & NDCG & \textbf{0.1742} & 0.1626& 0.1721& 0.1595\\
        
        \hline
        \multirow{4}{*}{Amazon} & \multirow{2}{*}{Elec} & Recall & \textbf{0.1276}& 0.0993& 0.1253& 0.1024 \\
        ~ & ~ & NDCG & \textbf{0.0628}& 0.0452& 0.0600& 0.056 \\
		\cline{2-7}
        ~ & \multirow{2}{*}{Movie} & Recall & \textbf{0.0577}& 0.0548& 0.0527& 0.0545\\
        ~ & ~ & NDCG & \textbf{0.0273} & 0.0258& 0.0248& 0.0252 \\
        \hline
  \end{tabular}
  }
\end{table}

\subsection{Ablation Studies}\label{ablation}

To evaluate the contributions of different components in our MDAP framework, we conducted ablation experiments on three datasets. We specifically examined the impact of the Gumbel-Softmax generator, the multi-view encoder, and the domain-specific gated network. The ablated models are as follows:

\begin{itemize}
    \item \textbf{MDAP-GS}: This variant removes the Gumbel-Softmax component and replaces it with a standard Softmax function. The Gumbel-Softmax technique is crucial for generating discrete soft assignments in a differentiable manner.
    \item \textbf{MDAP-MV}: This variant does not utilize multiple views, effectively reducing the number of views to 1. The multi-view encoder is designed to capture diverse aspects of user preferences.
    \item \textbf{MDAP-DG}: This variant removes the domain-specific gated network, and instead, combines all views by equally weighting them. The gated network is intended to adaptively combine embeddings from different views.
\end{itemize}

Table \ref{tab: ablation study} presents the results of our ablation study. Overall, the removal of each component results in a performance degradation across all datasets.

\textbf{Gumbel-Softmax Generator (MDAP-GS):} The ablation of the Gumbel-Softmax generator leads to a noticeable decline in performance. The Gumbel-Softmax technique enables the model to generate discrete soft assignments while maintaining differentiability, thus capturing more precise and diverse user preferences. The standard Softmax function fails to provide this granularity, leading to less effective feature representation.

\textbf{Multi-view Preference Encoder (MDAP-MV):} When the multi-view preference encoder is replaced with a single-view encoder, the performance drops significantly. This demonstrates the importance of capturing multiple aspects of user preferences through different views. The multi-view preference encoder effectively disentangles user preferences into multiple dimensions, improving the robustness and accuracy of recommendations.

\textbf{Domain-specific Gated Network (MDAP-DG):} The absence of the domain-specific gated network also results in a performance decrease. The domain-specific gated network's role is to adaptively combine the embeddings from different views, tailored for each domain. Simply averaging the embeddings without considering domain-specific characteristics does harm to the model's ability to make precise recommendations.

The ablation study highlights the significance of each component in the MDAP framework. The Gumbel-Softmax generator, multi-view encoder, and domain-specific gated network each contribute to the overall effectiveness of the model. The degradation in performance observed in the ablated versions confirms that these components are essential for capturing diverse user preferences and making accurate cross-domain recommendations.

\section{Conclusion}\label{conclusion}

In this paper, we proposed the MDAP framework, a novel multi-view disentangled and adaptive approach for cross-domain recommendation. Our comprehensive experimental evaluation, conducted across three public datasets, demonstrates the superior performance of MDAP compared to state-of-the-art single-domain, multi-task learning, and cross-domain recommendation methods. 

The key contributions of our work include the integration of Gumbel-Softmax for view selection, the utilization of a multi-view approach to capture diverse user preferences, and the introduction of a gated network to effectively combine multi-view preference information. Our ablation studies highlight the importance of each component within the MDAP framework, showing that removing any single module results in a noticeable performance decline.
In conclusion, the MDAP framework represents a significant step forward in the field of cross-domain recommendation systems, offering an adaptive and effective solution to leverage multi-view information for improving recommendation accuracy.

\paragraph{}
%
%
%
%




\bibliographystyle{splncs04}
\bibliography{content/References}

\end{document}